\documentclass[twocolumn,aps,showpacs]{revtex4}
\usepackage{graphicx}%
\newcommand{\ba}{\begin{eqnarray}}
\newcommand{\ea}{\end{eqnarray}}
\begin{document}
\title{ Quantum chaos and nuclear mass systematics}

\author{ Jorge G. Hirsch, V\'\i ctor Vel\'azquez and Alejandro Frank}
\affiliation{Instituto de Ciencias Nucleares,
Universidad Nacional Aut\'onoma de M\'exico, \\
Apartado Postal 70-543, 04510 M\'exico, D.F., M\'exico\\
E-mail: frank@nuclecu.unam.mx, hirsch@nuclecu.unam.mx, vic@nuclecu.unam.mx}

\begin{abstract}
The presence of quantum chaos in nuclear mass systematics is
analyzed by considering the differences between measured and
calculated nuclear masses as a time series described by the power
law $1/f^\alpha$. While for the liquid droplet model plus shell
corrections a quantum chaotic behavior  $\alpha \approx
1$ is found, errors in the microscopic mass formula have   
$\alpha \approx 0.5 $, closer to white noise. The chaotic behavior seems to arise
from many body effects not included in the mass formula. 
\end{abstract}

\maketitle

The importance of  an accurate knowledge of nuclear masses to understand diverse processes in nuclear physics and astrophysics is well known \cite{Rol88}. 
Though tremendous progress has been made in the challenging task of measuring the mass of exotic nuclei, theory is necessary to predict the mass of nuclei very far from stability \cite{Lunn03}. 
Understanding of the properties of complex nuclei in terms of the actual forces between nucleons is a basic question which at present nuclear is unable to answer \cite{Lunn03}. For this reason nuclear masses are predicted using phenomenological models.
The finite range droplet model (FRDM), which combines a macroscopic droplet with microscopic shell and pairing corrections \cite{Moll95}, has become the {\em de facto} standard for mass formulas. A microscopically inspired model
was successfully introduced by Duflo and Zuker (DZ)\cite{Duf94}. Along the
mean field model it is worth to mention the powerful Skyrme-Hartree-Fock  (HFB) approach \cite{Gor01}.
All these mass formulas can calculate and predict the masses (and often
other properties) of as many as 8979 nuclides \cite{Lunn03}. There is a
permanent search for  better theoretical models that reduce the
difference with the experimental masses and produce reliable
predictions for unstable nuclei. 
At present, the rms error for 2135 nuclei is 674 keV for HFB, 676 keV for FRDM, and 373 keV for DZ. The origin of the differences in rms errors between the models and the possibility of reducing them are the subject of the present investigation.

Recently, the problem of the mass deviations was analyzed from a new angle: in Ref. \cite{Boh02} the errors among experimental and calculated masses in \cite{Moll95} were interpreted in terms of two types of contributions. The first one was associated with a regular part, related to the underlying collective dynamics (droplet model), plus the shell energy correction, while the other was assumed to arise from some inherent dynamics, possibly higher order interactions among nucleons \cite{Boh02}, that lead to chaotic behavior. According to \cite{Abe02} the latter could be interpreted as remaining signals of the chaotic dynamics occurring at higher energies, whose magnitude suggests that we have already achieved (within a factor of 2) the maximum accessible precision in the calculation of the masses in mean-field theories \cite{Boh02}. 
It is relevant to ask if this chaotic limit can be confirmed by independent techniques and, if so, if this lower bound is valid also for mass calculations which explicitly include residual correlations, like DZ.

The presence of chaotic motion in nuclear systems has been firmly related with the statistics of high-lying energy levels  \cite{Wig65,Meh90}. Poisson distributions  of normalized spacings of successive nuclear or atomic excited levels with the same spin and parity correspond to integrable classical dynamics, while Wigner's statistics signal
chaotic motion in the corresponding classical regime \cite{Boh85}. Intermediate
situations are more difficult to assess.
Very recently a proposal has been made to treat the spectral fluctuations $\delta_n$ as discrete time series 
 \cite{Rel02}. Defining
\begin{equation}
\delta_n = \int_{-\infty}^{E_{n+1}} \tilde\rho(E) dE - n ,
\end{equation}
with $\tilde\rho(E)$ the mean level density which allows the mapping
to dimensionless levels with unitary average level density, and analyzing the energy 
fluctuations as a discrete time series, they found that nuclear power spectra behave like
$1/f$ noise, postulating that this might be a characteristic signature of generic quantum chaotic systems.
In the present work we implement this idea, using the $1/f$ spectral behavior as a test for the presence of chaos in nuclear mass errors.

In \cite{Fra03} a systematic study of nuclear
masses was carried out using the shell model, in an attempt
to clarify the nature of the errors. This was achieved by employing
realistic Hamiltonians with a small random component. In
\cite{Vel03} we have analyzed in detail the error distribution for
the mass formulas of M\"oller et al.\cite{Moll95} and found a conspicuous long
range regularity that manifests itself as a double peak in the
distribution of mass differences \cite{Vel03}. This striking
non-Gaussian distribution was found to be robust under a variety
of criteria. By assuming a simple sinusoidal correlation, we could
empirically substract these correlations and made the average
deviation diminish by nearly $15\%$ \cite{Hir03}.

In the present letter we carry out a study of the mass deviations
in the Finite Range Droplet Model (FRDM) of M\"oller et al. \cite{Moll95}, and
in the microscopically motivated mass formula of DZ \cite{Duf94,Duf95},
analyzing their correlations as time series. Two
different criteria are employed to organize the data, which render
similar and consistent power laws.

To map the mass error data, which depend on the charge $\cal Z$ and neutron number $\cal N$,
in term of variables with the maximum possible number
of nuclei along each chain, the following transformation is employed
\begin{eqnarray}
\tilde A = \hbox{Int} [ \sqrt{2} \,( {\cal N} \,sin \theta \,+\, {\cal Z}\, cos \theta ) ] , \nonumber \\
\tilde T_z = \hbox{Int} [ \sqrt{2}\, ( {\cal N} \, cos \theta \,-\, {\cal Z} \,sin \theta )].
\end{eqnarray}
Both $\tilde A$ and $\tilde T_z$ are, by construction, integer
numbers. To avoid introducing artificial noise, the data are {\em
softened} by the interpolation of mass errors for {\em unphysical}
values  of $\tilde T_z, \tilde A$, i.e. those with $\tilde T_z$
even and $\tilde A$ odd, or vice versa. This process is necessary
to eliminate the large number of zeroes which are induced by the
transformation, which create artificial high frequency noise in
the data.

 We found that the best orientation, in order to have
as many isotopes as possible with the same $\tilde T_z $, is
$\theta = 56 ^\circ$. With this transformation, e.g., there are
174 isotopes with $\tilde T_z = 0$.

\begin{figure}[h!]
  \begin{center}
    \includegraphics[width=8.0cm]{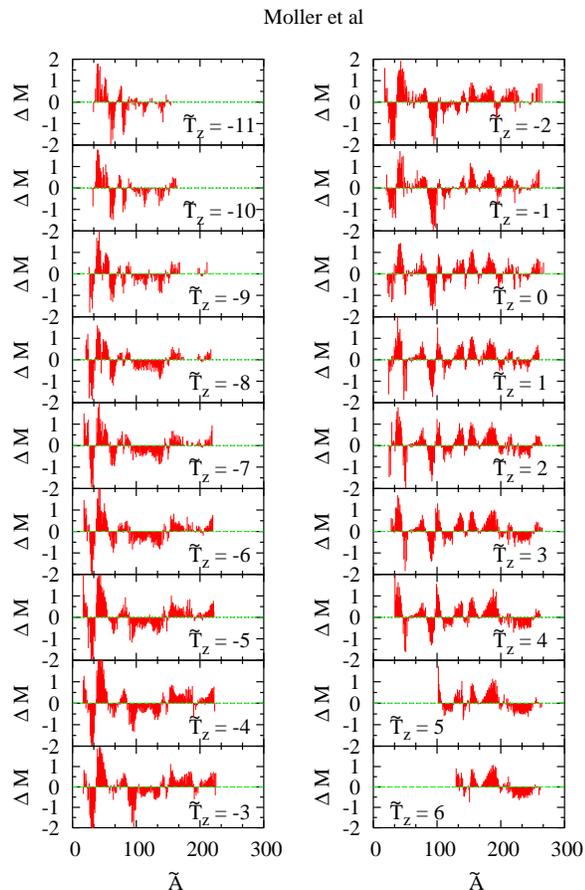}
  \end{center}
\caption{Mass differences from the FRDM calculations, in MeV, as functions
of $\tilde A$, for 18 $\tilde T_z$ values.}
\label{difmas-mn}
\end{figure}
Figure \ref{difmas-mn} displays the mass errors $\Delta M(\tilde
A) = M_{th}-M_{exp}$ for 18 values of $\tilde T_z$, from $\tilde
T_z = -11$ to 6 for the FRDM calculations. The
regularities seen in Fig. 2 of Ref. \cite{Hir03} as regions with
the same gray tone are seen here in the different plots, as
groupings of nuclei with similar positive or negative mass
differences, for the same $\tilde A$ region. Besides the two large
groups with positive and negative mass errors below $\tilde A=50$,
there are evident regions with negative errors close to $\tilde
A=100$, and with positive mass differences for $150< \tilde A
<200$.

\begin{figure}[h!]
  \begin{center}
    \includegraphics[width=8.0cm]{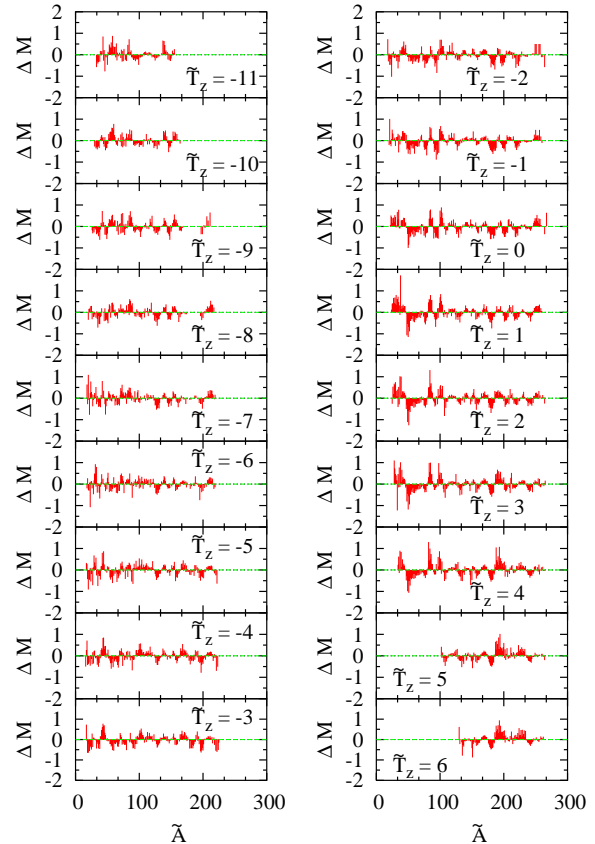}
  \end{center}
\caption{Mass differences from the Duflo and Zuker calculations, in MeV, as
functions of $\tilde A$, for 18 $\tilde T_z$ values.}
\label{difmas-dz}
\end{figure}
Figure \ref{difmas-dz} displays the mass errors for 18 values of $\tilde T_z$
for the DZ calculations. The deviations are manifestly smaller and
exhibit considerably less structure.

\begin{figure}[h!]
  \begin{center}
    \includegraphics[width=8.0cm]{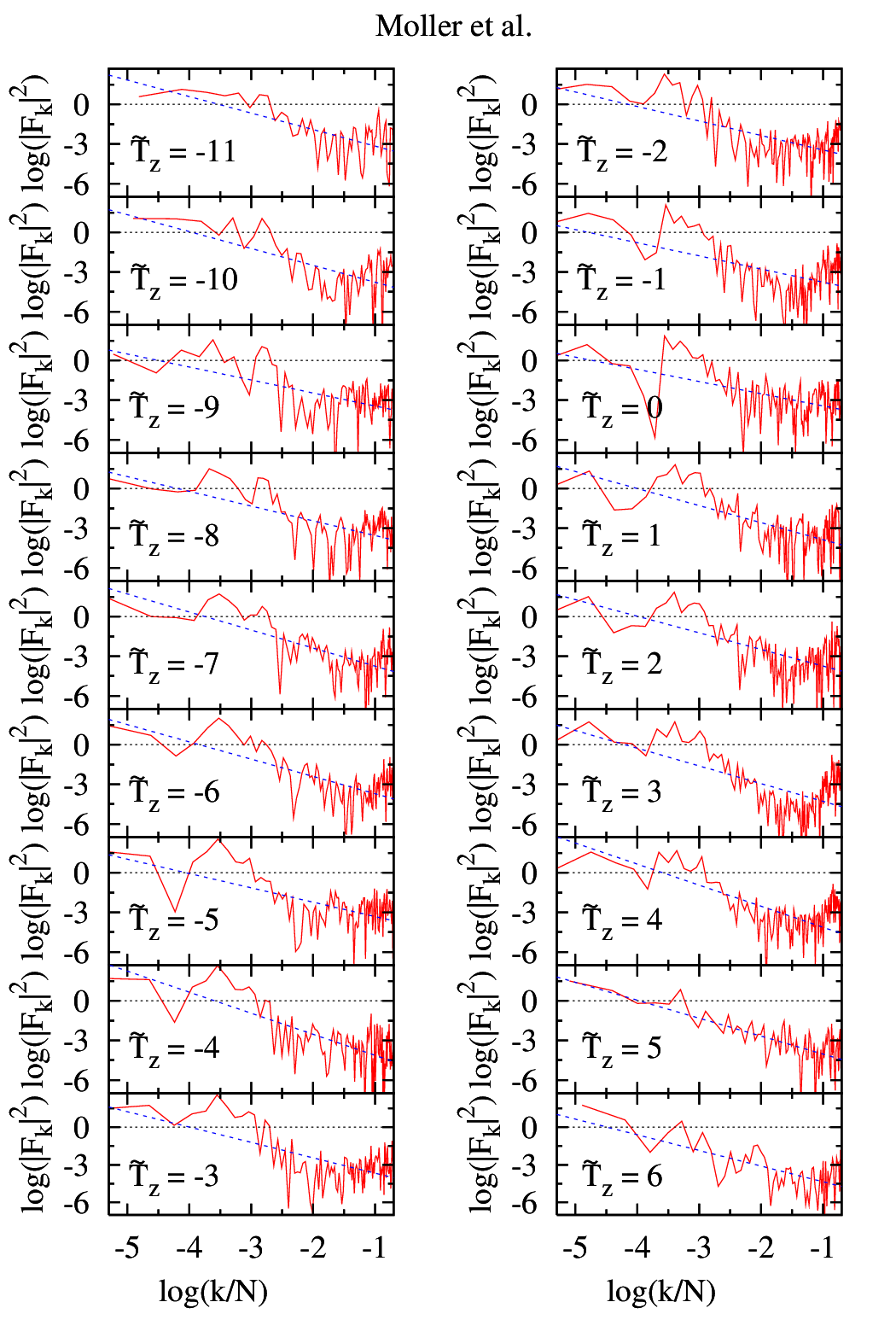}
  \end{center}
\caption{Logarithm of squared amplitudes of the Fourier transforms of the mass
differences, plotted as functions of the logarithm of the frequency, for 18 $\tilde T_z$
values, using the M\"oller at al data.}
\label{log-four-difmas56-mn}
\end{figure}

\begin{figure}[h!]
  \begin{center}
    \includegraphics[width=8.0cm]{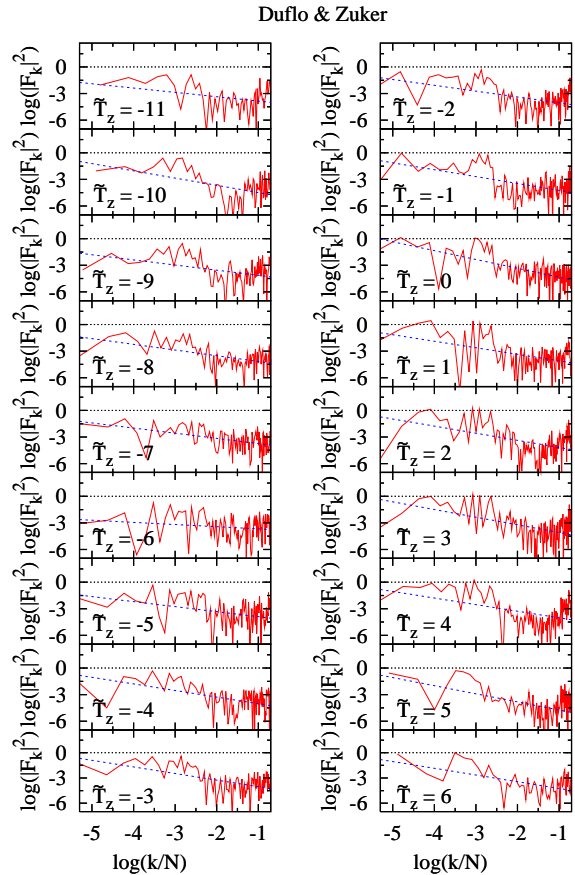}
  \end{center}
\caption{Logarithm of squared amplitudes of the Fourier transforms of the mass
differences, plotted as functions of the logarithm of the frequency, for 18 $\tilde T_z$
values, using the Duflo and Zuker data.}
\label{log-four-difmas56-dz}
\end{figure}

The discrete Fourier $F_k$ transforms are calculated as
\begin{equation}
F_k = {\frac {1} {\sqrt{ N}}} \sum_j {\frac {\Delta M(j)}{\gamma}} \,  \hbox{exp}
\left({\frac {-2 \pi i j k} { N}}\right) ,
\end{equation}
where $N$ is the number of mass differences $\Delta M$ in a given series.
The parameter $\gamma$ makes $F_k$ dimensionless. Given that it only affects 
the global scale of the Fourier amplitudes, we made the simple selection
$\gamma = 1$ MeV.
The Fourier amplitudes are plotted as functions of the logarithm of the frequency $f = k/{N}$
for the FRDM data in Fig. \ref{log-four-difmas56-mn} and for
the Duflo and Zuker data in Fig.  \ref{log-four-difmas56-dz}, using a log-log scale.

These plots have some remarkable features. As expected from a $1/k^\alpha$ power law,
low frequencies have the larger amplitudes. In most of the plots the largest Fourier
amplitudes are those with frequencies between 0.3 and 0.4 (log f = -4 to -3),
indicating that oscillations with periods $\Delta  \tilde A \approx 20-50$
are dominant. This is consistent with the fitted frequencies found in Ref. \cite{Hir03}.
A slight rise of the amplitudes at the larger frequencies (f $\approx$ 0.5, 
$\Delta  \tilde A \approx 2$)
can be seen in many plots. They represent strong fluctuations between some nuclei and their closest neighbors.

The Fourier amplitudes are consistently smaller for the DZ data,
which have also a Gaussian-like distribution of the mass
differences \cite{Vel03}. This indicates that
the FRDM mass differences have stronger correlations,
which are precisely the ones removed in Ref. \cite{Hir03}.

The straight lines correspond to the best fitted slopes, in the log-log plots, of the power spectra, that is, the squared Fourier amplitudes against the frequency.
While the fluctuations are large, and the number of nuclei
included in each chain range from a few dozen to almost two
hundred, the results are striking and correlated with the
recently proposed universal features of quantum chaos
\cite{Rel02}. 

For the 18 chains listed, the slopes are
\begin{equation}
\alpha^{(1)}_{FRDM} = -1.18 \pm  0.17, \; ~~~\alpha^{(1)}_{DZ} = -0.67 \pm 0.16.
\end{equation}
They fluctuate around -1.2  in the FRDM data and around -0.7 for the deviations found by DZ.
These slopes convey our main result. The former is consistent with a
frequency dependence of  $f^{-1}$ characteristic of quantum chaos while the 
latter suggest a tendency towards a more random behavior characteristic of white noise.

\begin{figure}[h!]
  \begin{center}
    \includegraphics[width=6.0cm,angle=270]{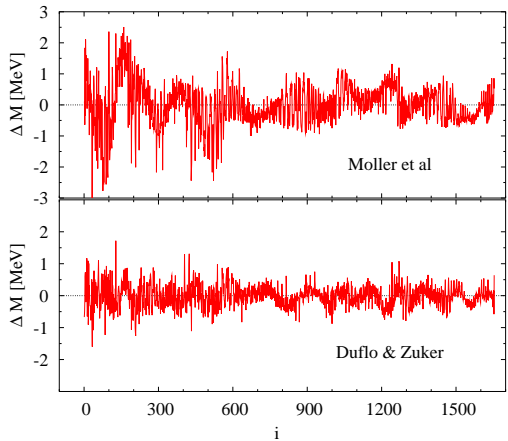}
  \end{center}
\caption{Mass differences plotted as an ordered list, taken from FRDM (top),
and from Duflo and Zuker (bottom).}
\label{difmas-l}
\end{figure}

An alternative way to organize the 1654 nuclei with measured
masses is to order them in a {\em bustrofedon} single list \cite{Hir03}, numbered in increasing
order. To avoid jumps, we have ordered the nuclei with even A
following the increase in N-Z, and those nuclei with odd A
starting from the largest value of N-Z, and going on in decreasing
order. Fig. \ref{difmas-l} exhibits the mass differences plotted
against the order number, from 1 to 1654, taken from M\"oller et
al (top)
 and from DZ (bottom).
The presence of strong correlations in the M\"oller at al mass
differences is apparent from the plot. Regions with large positive
or negative errors are clearly seen. In the data of Duflo and
Zuker the distribution of errors is closer to the horizontal axis,
and the correlations are less pronounced, although not completely
absent.

\begin{figure}[h!]
  \begin{center}
    \includegraphics[width=6.0cm,angle=270]{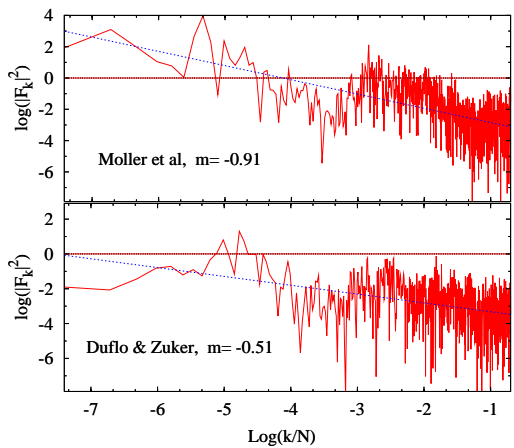}
  \end{center}
\caption{Log-log plot of the squared amplitudes of
the Fourier transforms of the mass differences, as functions of
the order parameter(top). Data from FRDM (top) and from
Duflo and Zucker (bottom).} \label{fourier-l}
\end{figure}

The ordering provides a single-valued function,
whose Fourier transform can be calculated. The squared amplitudes are presented
in Fig. \ref{fourier-l}.
The slopes are 
\begin{equation}
\alpha^{(2)}_{FRDM} = -0.91 \pm  0.05, \; ~~~\alpha^{(2)}_{DZ} = -0.51 \pm 0.05,
\end{equation}
 for the FRDM and DZ mass differences. 

While this ordering is quite different from the
$\tilde A$ chains, the slopes are very similar.

To understand the possible origin of these spectral distributions,
it is worth recalling that, while the FRDM calculations
involve a liquid droplet model plus mean field corrections,
including deformed single particle energies through the Strutinsky
method and pairing \cite{Moll95}, the DZ calculations depend on
the number of valence proton and neutron particles and holes,
including quadratic effects motivated by the microscopic
Hamiltonian \cite{Zuk94}.
The present results show that the DZ formalism produces patterns that are locally smooth approximations to the data, and therefore give some information on the intrinsic nature of the data fluctuations.

We arrive at the conclusion that the chaoticity discussed in
\cite{Boh02}, according to the criteria  put forward in \cite{Rel02}, seems indeed to be present in the deviations induced by calculations using the M\"oller et al. liquid droplet mass
formula, while it tends to diminish in the microscopically motivated
calculations of Duflo and Zuker. 
 While for the liquid droplet model plus shell
corrections a quantum chaotic behavior  $\alpha \approx
1$ is found, errors in the microscopic mass formula have   
$\alpha \approx 0.5 $, closer to white noise.
Given that both models attempt
to describe the same set of experimental masses, our analysis
suggests that quantum fluctuations in the mass differences arising
from substraction of the regular behavior provided by the liquid droplet model plus shell corrections, may have their origin in an incomplete consideration of many body quantum correlations, which are partially included in the calculations of Duflo and Zuker.
 This interpretation would imply that it
is in principle possible to reduce the limits in accuracy mentioned in
\cite{Abe02} for the calculation of nuclear masses. 
It remains to be seen whether a robust picture of the coexistence of regular and chaotic motion emerge from these studies and whether a quantitative means to evaluate their relative importance can be formulated.

Acknowledgements: Relevant comments by R. Bijker, O. Bohigas, J. Dukelsky, J. Flores, J.M. Gomez, P. Leboeuf, R. Molina, S. Pittel, A. Raga, P. van Isacker, and A. Zuker are gratefully acknowledged.
This work was supported in part by Conacyt, M\'exico.


\begin{thebibliography}{aa}
\bibitem{Rol88} C.E. Rolfs and W.S. Rodney, {\em Cauldrons in the Cosmos},
University of Chicago Press (1988).
\bibitem{Lunn03} D.Lunney, J.M. Pearson, and C. Thibault, Rev. Mod. Phys. {75}, 1021 (2003).
\bibitem{Moll95} P. M\"oller, J.R. Nix, W.D. Myers, W.J. Swiatecki,
At. Data Nucl. Data Tables {\bf 59}, 185 (1995).
\bibitem{Duf94} J. Duflo, Nucl. Phys. {\bf A 576}, 29 (1994);
J. Duflo and A. P. Zuker, Phys. Rev. {\bf C 52}, R23 (1995).
\bibitem{Gor01} S. Goriely, F. Tondeur, and J.M. Pearson, Atom. Data Nucl. Data
Tables {\bf 77}, 311 (2001).
\bibitem{Boh02} O. Bohigas, P. Leboeuf, Phys. Rev. Lett. {\bf 88}, 92502 (2002).
\bibitem{Abe02} S. \AA berg, Nature {\bf 417}, 499 (2002).
\bibitem{Meh90} M.L. Mehta, {\em Random Matrices}, Academic Press, London
(1990).
%\bibitem{Pac02} A. de Pace, A. Molinari, arXiv:nucl-th/0209020.
\bibitem{Wig65} E.P. Wigner, {\em Statistical Theories of Spectra: Fluctuations},
Academic, New York (1965).
\bibitem{Boh85} O. Bohigas, R.U. Haq, A. Pandey, Phys. Rev. Lett. {\bf 54}, 1645 (1985),
and references therein.
\bibitem{Rel02}  A. Rela\~no, J.M.G. G\'omez, R.A. Molina, J. Retamosa and E. Faleiro,
Phys. Rev. Lett. {\bf 89} (2002) 244102.
\bibitem{Fra03} V\'\i ctor Vel\'azquez, Jorge G. Hirsch, and Alejandro Frank,
Rev. Mex. F\'\i s. {\bf 49} S. 4 (2003) 34.
\bibitem{Vel03} V\'\i ctor Vel\'azquez, Alejandro Frank, and Jorge G. Hirsch,
%Proc. of the workshop 
{\em Computational and Group Theoretical Methods in Nuclear Physics}, 
%Playa del Carmen, Mexico, February 18 - 22, 2003,
Eds. J. Escher, O. Casta\~nos, J.G. Hirsch, S. Pittel, G. Stoitcheva,
World Scientific, Singapore (2004) 51.
\bibitem{Hir03} Jorge G. Hirsch, Alejandro Frank, and V\'\i ctor Vel\'azquez, 
Phys. Rev. {\bf C 69}, 37304 (2004).
\bibitem{Duf95} http://csnwww.in2p3.fr/AMDC/theory/du\_zu\_28.feb95
\bibitem{Zuk94}  A. P. Zuker, Nucl. Phys. {\bf A 576}, 65 (1994).
\end{thebibliography}
\end{document}